\documentclass[letterpaper]{aipproc}

\layoutstyle{6x9}

\def\be{\begin{equation}}
\def\ee{\end{equation}}
\def\bea{\begin{eqnarray}} 
\def\eea{\end{eqnarray}}

\def\Journal#1#2#3#4{{#1} {\bf #2}, #3 (#4).}
\def\ANP {\it Advances in Nucl. Phys.}
\def\ANN {\it Ann. Phys. (N.Y.)}

\def\EPJA{{\it Eur. Phys. J} A}

\def\NCA{\it Nuovo Cimento}

\def\NPA{{\it Nucl. Phys.} A}

\def\PLB{{\it Phys. Lett.}  B}

\def\PR{\it Phys. Rev.}
\def\PRL{\it Phys. Rev. Lett.}

\def\PRC{{\it Phys. Rev.} C}

\begin{document}    

\rightline{\parbox{1.5in}{\leftline{JLAB-PHY-01-23}
                \leftline{WM-01-112}
			             }}

\title
[deuteron review]
{The deuteron: a mini-review}


\author{\underline{Franz Gross}$^{*}$}{
  address={Department of Physics, College of William and Mary, 
   Williamsburg, VA 23187, USA},
  address={Jefferson Laboratory, 12000
    Jefferson Avenue, Newport News, VA 23606, USA} }
 
\author{R.~Gilman$^{**}$}{
  address={Rutgers University, 136 Frelinghuysen Rd,
Piscataway, NJ 08855, USA},
  address={Jefferson Laboratory, 12000
    Jefferson Avenue, Newport News, VA 23606, USA} }

\begin{abstract}
We review\thanks{This talk is a shorter version of a review being 
prepared for {\it Journal of Physics G\/} \cite{GG01}.} some recent 
results for elastic electron deuteron  scattering (deuteron form 
factors) and photodisintegration of the deuteron, with emphasis on the 
recent high energy data from Jefferson Laboratory (JLab). 
\end{abstract}

\date{\today} 
  
\maketitle

\section{Deuteron wave functions and form factors }

\begin{figure}
\caption{Momentum space wave functions for five models 
mentioned  in the text: AV18 (solid), Paris (long
dashed), CD Bonn (short dashed), IIB (short dot-dashed), and
W16 (long dot-dashed)  The wave
functions in the right panel have been divided by scaling 
functions for easy comparison (see Ref.\ [1] for details).}
\label{uwp}
\includegraphics[height=.33\textheight]{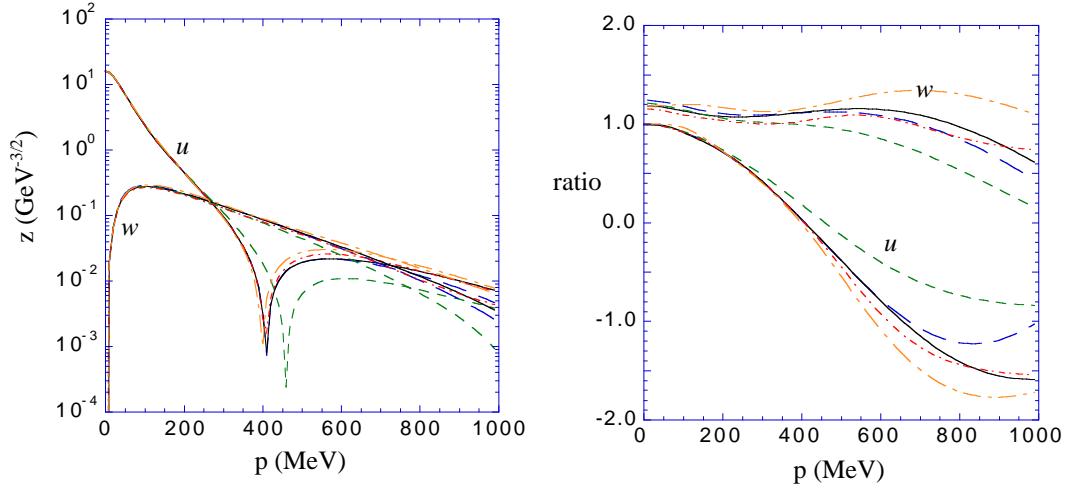}
\end{figure}

Calculations of deuteron form factors and photo and 
electrodisintegration to the $NN$ final state require a deuteron wave 
function.  The best nonrelativistic wave functions are calculated from 
the Schr\"odinger equation using a potential adjusted to fit the $NN$ 
scattering data for lab energies from 0 to 350 MeV.  The quality of 
realistic potentials have improved steadily, and now the best 
potentials give fits to the $NN$ database with a $\chi^2$/d.o.f 
$\simeq1$.  The Paris potential \cite{Paris} was among the first 
potentials to be determined from such realistic fits, and it has since 
been replaced by the Argonne V18 potential (denoted by AV18) 
\cite{AV18}, the Nijmegen potentials \cite{Nijmegen}, and most recently 
by the CD Bonn potentials \cite {CDBonn,CDBonn2000}.  The momentum 
space $S$ and $D$ state wave functions determined from three of these 
models and two relativistic models (Model IIB \cite{gvoh92} and Model 
W16, one of a family of models with varying amounts of off-shell sigma 
coupling introduced in connection with relativistic calculations of the 
triton binding energy described in Ref.~\cite{sg97}) are shown in 
Fig.~\ref{uwp}. The figure shows that the  $S$ and $D$-state components 
of all of these models are  almost identical (i.e.variations of less 
than 10\%) for momenta below about 400 MeV, and vary by less than a 
factor of 2 as the momenta reaches 1 GeV. 

Elastic electron--deuteron scattering is described in the one-photon 
exchange approximation by three deuteron form factors 
\cite{G65,acg80,donn86}.   In its most general form, the relativistic 
deuteron current can be written \cite{G65,HJ62} 
\begin{eqnarray}
-\left<d'|J^\mu|d\right>=\left\{G_1(Q^2)\;
[\xi'^*\cdot\xi]-G_3(Q^2)\frac{(\xi'^*\cdot q)(\xi\cdot q)}{2m_d^2}\right\}
\,(d^\mu+d'^\mu)  \nonumber\\
 + G_M(Q^2)\;[\xi^\mu(\xi'^*\cdot q)
-\xi'^{*\mu}(\xi\cdot q)]\, ,\label{deutcurrent}
\end{eqnarray} 
where $\xi, d$ ($\xi', d'$) are the polarization and momentum vectors 
of the incoming (outgoing) deuterons, and the form factors $G_i(Q^2)$, 
$i=1-3$, are all functions  of $Q^2=-q^2$, the square of the {\it 
four\/}-momentum transferred by the electron, with $q=d'-d$.  In 
practice, $G_1$ and $G_3$ are replaced by a more physical choice of 
form factors 
\begin{eqnarray}
&&G_C=G_1 +\frac{2}{3}\,\eta\,G_Q\nonumber\\ 
&&G_Q=G_1-G_M+(1+\eta)G_3\, ,
\end{eqnarray}
with $\eta=Q^2/4m_d^2$.  At $Q^2=0$, the form factors $G_C$, $G_M$, and 
$G_Q$ give the charge, magnetic and quadrupole moments of the deuteron 
\begin{eqnarray}
G_C(0) =& 1 &\qquad ({\rm in\ units\ of}\, e )\nonumber\\ 
G_Q(0) =& Q_d &\qquad ({\rm in\ units\ of}\, e/m_d^2)\nonumber\\
G_M(0) =&\mu_d &\qquad ({\rm in\ units\ of}\, e/2m_d) \, .     
\end{eqnarray}

The structure functions $A$ and $B$, and the polarization transfer 
coefficient $T_{20}$ depend on the three electromagnetic form factors 
\begin{eqnarray}
&&A(Q^2) = G_C^2(Q^2) + {{8}\over{9}} \eta^2 G_Q^2(Q^2) + 
 {{2}\over{3}} \eta G_M^2(Q^2)\nonumber\\
&&B(Q^2) = {{4}\over{3}} \eta(1+\eta) G_M^2(Q^2)\nonumber\\
&&\tilde{T}_{20} = - \sqrt{2}\; { {y(2+y)}\over{1 + 2 y^2} }\, ,
\label{AandB}
\end{eqnarray}
where $y = 2 \eta G_Q / 3 G_C$, and $\tilde{T}_{20}$ is
$T_{20}$ with the magnetic contributions removed.  

\begin{figure}
\caption{The structure function $A$ for five nonrelativistic
models using the MMD nucleon form factors.  The models are
labeled as in Fig.~1.  The right panel shows data and
models divided by a ``fit'' described in Ref.~[1].  The data
are fully referenced in [1].}
\includegraphics[height=.31\textheight]{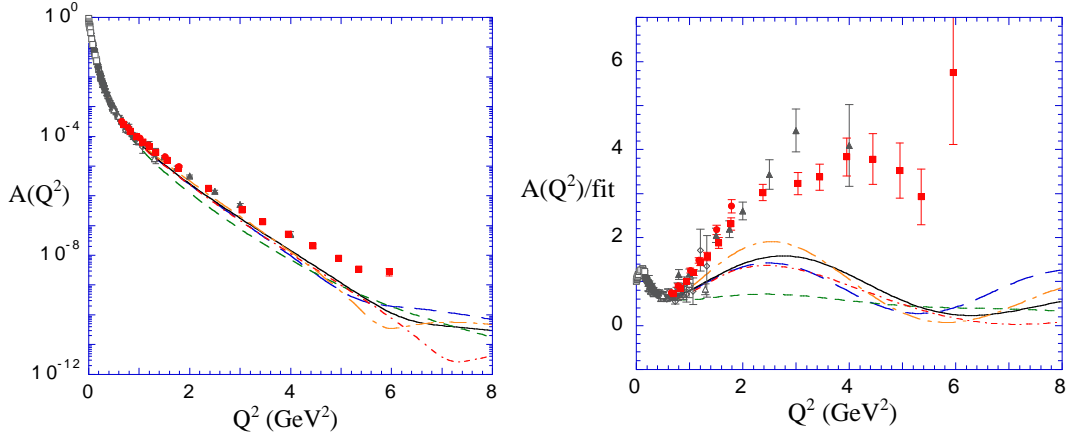}
\label{AhighQ2}
\end{figure}

\subsection{Comparison of nonrelativistic theory to data}

In the nonrelativistic theory, {\it without exchange currents or 
$(v/c)^2$ corrections\/}, the deuteron form factors are 
\begin{eqnarray}
G_C  &=& G_{E}^sD_C \nonumber\\
G_Q  &=& G_{E}^sD_Q \nonumber\\
G_M  &=& \frac{m_d}{2m_p}\Bigl[G_{M}^s D_M + G_{E}^s D_E\Bigr]\, ,
\label{body}
\end{eqnarray}  
where $G^s_E$ and $G^s_M$ are the {\it nucleon isoscalar\/} form
factors, and the $D$s are the {\it body\/} form factors.  All
are functions of $Q^2$.  Hence the study of deuteron form
factors is complicated by the fact  that they are a {\it
product\/} of the {\it nucleon isoscalar\/} form factors and
{\it body\/} form factors.  The dependence of the deuteron form 
factors on older models of the nucleon form factors is well
discussed in Ref.~\cite{GVO01}. A year ago the  model of
Mergell, Meissner and Drechsel \cite{MMD} (referred to as MMD)
gave a good fit, and could have been adopted as a standard
(the new $G_{Ep}/G_{Mp}$ data from JLab \cite{jones99} may
change this view). The high
$Q^2$ data for
$A$ provide the most stringent test.   In Fig.~\ref{AhighQ2} we
compare the data for
$A$ with nonrelativistic calculations using the five
nonrelativistic wave functions shown in Fig.~\ref{uwp}.  The
calculations use Eq.~(\ref{body}) with MMD isoscalar nucleon
form factors and nonrelativistic body form factors.  In the
right panel the data and models have been divided by the ``fit''
described in Ref.~[1].

It is easy to see that the nonrelativistic models {\it are a 
factor 4 to 8 smaller than the data for $Q^2>
2$ GeV$^2$\/}.  Furthermore,
since the  difference between different deuteron models is
substantially smaller than this discrepancy, it is unlikely 
that any {\it realistic\/} nonrelativistic model can be found
that will agree with the data.  If the nucleon isoscalar charge
form factor were larger than the MMD model by a factor of 2 to
3 it might explain the data, but this is also unlikely since
the variation between nucleon form factor models is
substantially smaller than this.  We are forced to conclude
that these high $Q^2$ measurements {\it cannot be explained by
nonrelativistic physics and present very strong evidence for
the presence of interaction currents, relativistic effects, or
possibly new physics.\/}

\subsection{Relativistic calculations and new physics} 

\begin{figure}
\caption{Exchange currents that might play a role in meson
theories.  (a) Large $I=1$ $\pi, \rho,$ and $\Delta$ currents
that do not contribute to the deuteron form factors, and (b)
possible $I=0$ currents that are identically zero.  The
currents that do contribute to the deuteron form factors are
shown in the second row: (c) ``pair'' currents from nucleon
$Z$-graphs; (d) ``recoil'' corrections; (e) two pion exchange
(TPE) currents; and (f) the famous $\rho\pi\gamma$ exchange
current.}
\label{MEC}
\includegraphics[height=.3\textheight]{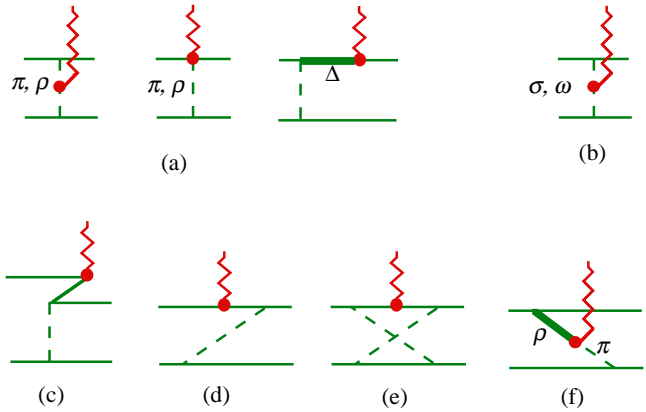}
\label{MEC}
\end{figure}

The differences between the data and the nonrelativistic theory
can only be explained by a combination of the following effects

\begin{itemize}
\item interaction (or meson exchange) currents;
\item relativistic effects; or
\item new (quark) physics.
\end{itemize}
The only possibilities excluded from this list are
variations in models of the nucleon form factors, or model dependence
of the deuteron wave functions.  Previously we have argued that
{\it neither\/} the current uncertainty in our knowledge of the
nucleon form factors, {\it nor\/} the model dependence of the
nonrelativistic deuteron wave functions is sufficient to provide
an explanation for the discrepancies.

Possible interaction currents that might account for the discrepancy 
are shown in Fig.~\ref{MEC}.  Because the deuteron is an isoscalar 
system, the familiar large $I=1$ exchange currents are ``filtered'' out 
and only $I=0$ exchange currents can contribute to the form factors. 
The $I=0$ currents tend to be smaller and of a more subtle origin.  The 
nucleon $Z$-graphs, Fig.~\ref{MEC}c, and the recoil corrections, 
Fig.~\ref{MEC}d, are both of relativistic origin.  (The recoil graphs 
will give a large, incorrect answer unless they are renormalized 
\cite{Gr91,JL75,TH73}.)  The two-meson exchange currents should be 
omitted unless the force also contains TPE forces.  The famous 
$\rho\pi\gamma$ exchange current is very sensitive to the choice of 
$\rho\pi\gamma$ form factor, which is hard to estimate and could easily 
be a placeholder for new physics arising from quark degrees of 
freedom. 
  
In most calculations based on meson theory, the two pion exchange (TPE) 
forces and currents are excluded, and, except for the $\rho\pi\gamma$ 
current (which we will regard as new physics), the exchange currents 
are of relativistic origin.  Additional relativistic effects arise from 
boosts of the wave functions, the currents, and the potentials, which 
can be calculated in closed form or expanded in powers of $(v/c)^2$, 
depended on the method used.  At low $Q^2$ calculations may be done 
using effective field theories in which a small parameter is 
identified, and the most general (i.e. exact) theory is expanded in a 
power series in this small parameter.  In these calculations, an 
expansion of relativistic effects in a power series in $(v/c)^2$ is 
automatically included.   {\it Hence\/}, with the exception of 
effective field theories, {\it any improvement on nonrelativistic 
theory using nucleon degrees of freedom leads us to relativistic 
theory.\/} 

Alternatively, one may seek to explain the discrepancy using quark 
degrees of freedom (new physics).  When two nucleons overlap, their 
quarks can intermingle, leading to the creation of new $NN$ channels 
with different quantum numbers (states with nucleon isobars, or even, 
perhaps, so-called ``hidden color'' states).  These models require that 
assumptions be made about the behavior of QCD in the nonperturbative 
domain, and are difficult to construct, motivate, and constrain.  At 
very high momentum transfers it may be possible to estimate the 
interactions using perturbative QCD (pQCD).  Very little has been done 
using other approaches firmly  based in QCD, such as lattice gauge 
theory or Skrymions.

\begin{figure}
\caption{The relativistic decision tree discussed in the text.}
\includegraphics[height=.3\textheight]{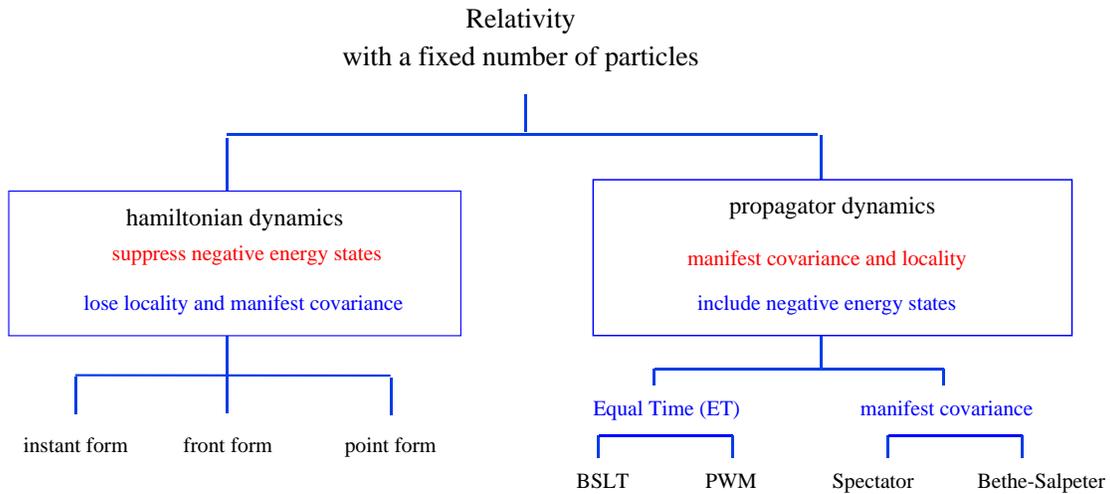}
\label{decision}
\end{figure}

In deciding which relativistic method to use, it is first necessary to 
decide whether or not to allow {\it antiparticle, or negative energy\/} 
nucleons to propagate as part of the virtual intermediate state.  Since 
nucleons are heavy and composite, so that their antiparticle states are 
very far from the region of interest, some physicists believe that 
intermediate states should be built only from positive energy nucleons, 
and that all negative energy effects (if any) should be included in the 
interaction.  These methods are referred to collectively as {\it 
hamiltonian dynamics\/} and are represented by the left hand branch 
shown in Fig.~\ref{decision}. Unfortunately, it turns out that this 
choice precludes the possibility of retaining the properties of 
locality and manifest covariance enjoyed by field theory.    
Alternatively, in order to keep the locality and manifest covariance of 
the original field theory, other physicists are willing to allow 
negative energy states into the propagators. These methods, represented 
by the right-hand branch of the figure, are referred to collectively as 
{\it propagator dynamics\/}.  However, including negative energy states 
tends to make calculations technically more difficult and harder to 
interpret physically, and those who advocate the use of hamiltonian 
dynamics do not believe the advantages of exact covariance justify the 
work it requires.  

Unfortunately, these two methods are so fundamentally different that  
many physicists do not realize that the limitations of one may not 
apply to the other.  For example, for some choices of propagator 
dynamics all 10 of the generators of the Poincar\'e group will depend 
only on the kinematics, and the Poincar\'e transformations of {\it all 
amplitudes can be done exactly\/}.  With hamiltonian dynamics this is 
not the case; some of the 10 generators must  depend on the 
interaction, and transformation of matrix elements under these 
``dynamical'' transformations must be calculated. Comparison of the two 
methods is therefore very difficult; the language and issues of each 
are very different and one can be easily misled by the different 
appearance of the results.

\subsection{Comparison of relativistic calculations with data}
 
\begin{figure}
\caption{The structure function $A$ for the eight
models discussed in the text. Left panels show the propagator
and instant-form results: FSR (solid line), VOG in RIA
approximation (long dashed line), ARW
(short dashed line), and PWM (dotted line).  Right panels
show the front-form CK (long dashed line) and LPS (short
dashed line), the point-form AKP (medium dashed line) and the
quark model calculation DB (solid line).  In every case the
calculations have been divided by a scaling function given
in Ref.~[1].    The data are labeled as in Fig~2.}
\includegraphics[height=.58\textheight]{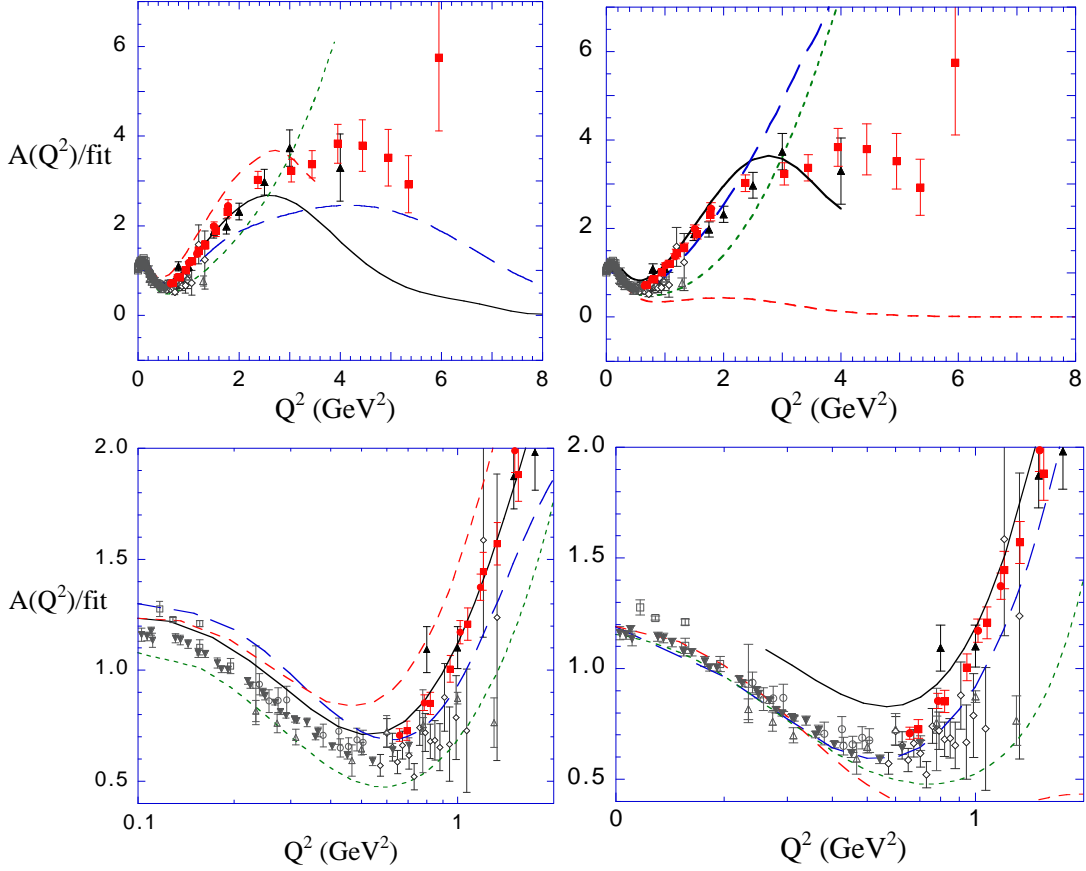}
\label{AlowQ27}
\end{figure}

\begin{figure}
\caption{The structure functions $A$, $B$, and $T_{20}$ for
the eight models discussed in the text.  VOG full
calculation (CIA plus $\rho\pi\gamma$ -- solid line); VOG in
RIA (long dashed line); FSR (medium dashed line); ARW (short
dashed line); DB (widely spaced dotted line); CK (long
dot-dashed line); AKP (short dot-dashed line); PWM (dashed
double-dotted line), and LPS (dotted line).   The data
are as labeled in Fig.~2.}
\includegraphics[height=.75\textheight]{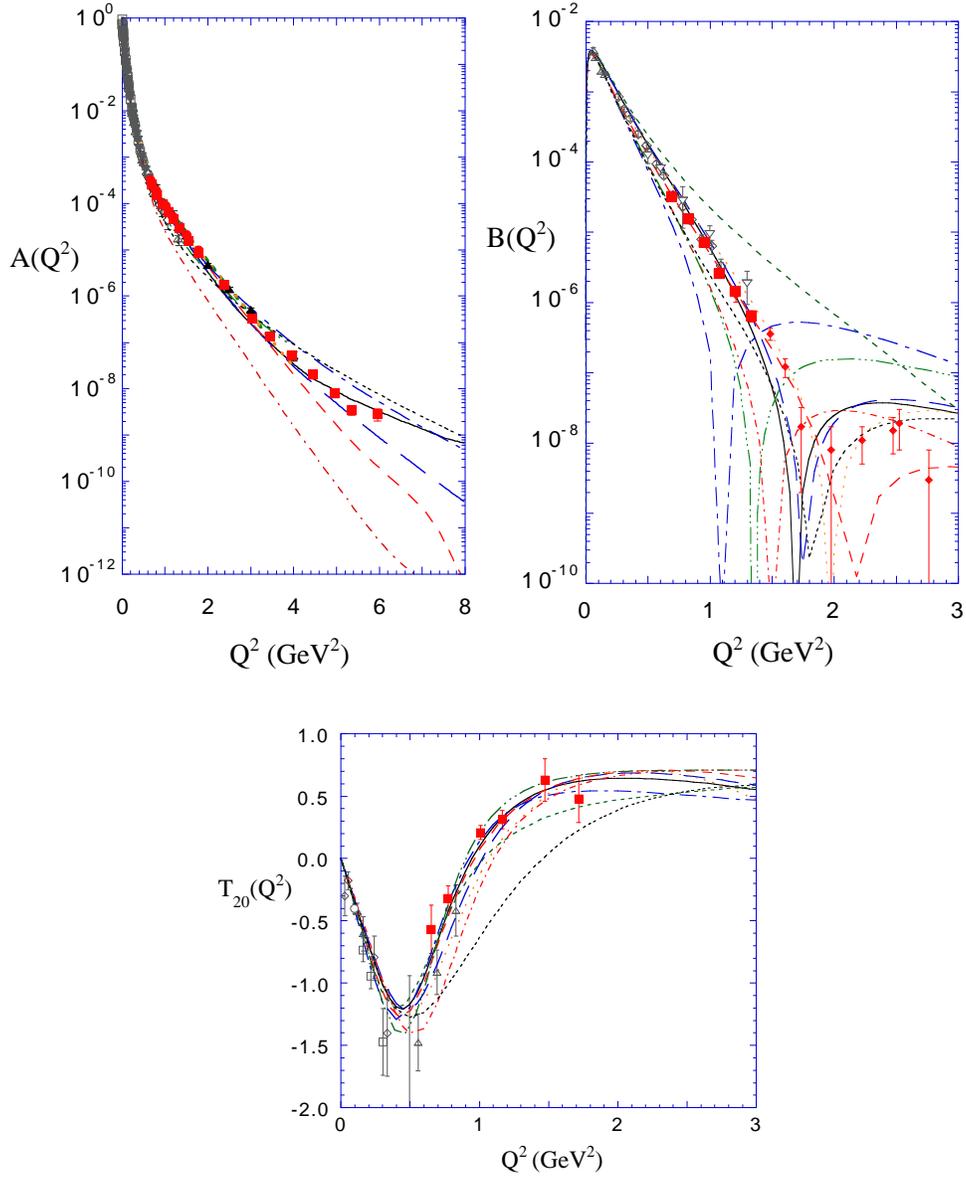}
\label{fig:RABT20}
\end{figure}

The high $Q^2$ predictions for 7 relativistic models using $NN$ degrees 
of freedom and one quark cluster model are shown in 
Figs.~\ref{AlowQ27}--\ref{fig:RABT20}.  They include 

\begin{itemize}
\item two propagator calculations: VGO \cite{vdg95}
(using the Spectator equation), and PWM \cite{PW} (using
the modified Mandelzweig-Wallace equation);

\item two instant-form calculations: FSR \cite{FS01}
(without a $v/c$ expansion) and ARW \cite{ARW00} (using a
$v/c$ expansion);

\item two front-form calculations: CK \cite{CK99} (with
the light front retained as an unphysical degree of
freedom)  and LPS \cite{LPS00} (using a specially
constructed current operator);

\item a point-form calcualtion: AKP \cite{AKP01}; and
  
\item a quark model calculation: DB \cite{DB88}.

\end{itemize}

The model dependence of the eight calculations is large. Figure 
\ref{AlowQ27} shows the predictions for $A(Q^2)$. In these figures we 
have intentionally left out the model dependent $\rho\pi\gamma$ 
exchange current from all of the calculations.  All of the models 
except the AKP point-form calculation give a reasonable description of 
$A$ out to $Q^2\sim 3$ GeV$^2$, beyond which they begin to depart 
strongly from each other and the data.  Taking into account that the 
$\rho\pi\gamma$ exchange current {\it could be added to any of these 
models, and that this contribution tends to increase A above $Q^2\sim 
3$ GeV$^2$\/}, four models seem to have the right general behavior: the 
VOG, FSR, ARW and the quark model of DB (but there are no results for 
this model beyond $Q^2=4$ GeV$^2$).  None of these models fit the data 
without a $\rho\pi\gamma$ exchange current, and all models would be 
improved by adding such a current contribution, {\it showing that there 
is some evidence for new physics at high $Q^2$\/}.  Ironically, none of 
the models favored by the high $Q^2$ data does as well at low $Q^2$ as 
the three ``unfavored'' models shown in the right panels (unless the 
Platchkov \cite{plat90} data are systematically too low).

Finally, Fig.~\ref{fig:RABT20} shows the predictions for the structure 
functions $A$, $B$, and $T_{20}$ for the eight models.  The LPS 
calculation shows a large discrepancy with the $T_{20}$ data, but the 
most striking feature of these plots is the {\it large model 
dependence\/} of the predictions for $B(Q^2)$.  The magnetic structure 
function provides the most stringent test of the models, and the 
predictions are comparatively free of the $\rho\pi\gamma$ exchange 
current (which gives only a small contribution to $B$).  Examination of 
the figure shows that the $B$ predictions of the PWM, ARW, AKP, CK 
models fare the worst.  In all, taking the predictions for the three 
structure functions together, the best results are obtained with the 
FSR, VOG, and DB models.

\subsection{Conclusions (deuteron form factors)}
\label{conclusions} 

What have we learned from our measurements of the deuteron form 
factors?  Our comparison of theory and experiment leads to the 
following conclusions: 

\begin{itemize} 

\item Nonrelativistic quantum mechanics (without exchange 
currents  or relativistic effects) is ruled out by the $A(Q^2)$
data at high $Q^2$.  Reasonable variations in nucleon form
factors or uncertainties in the nonrelativistic wave functions
cannot  remove the discrepancies.
 
\item In approaches using $NN$ degrees of freedom only,
relativistic effects (or $I=0$ meson exchange currents) 
could be large enough to explain the data. 
 
\item Some models that include  relativistic effects
(or meson exchange currents) and use $NN$ degrees of
freedom with realistic  forces are close to
the data.  None are entirely satisfactory.

\item The model dependence of relativistic effects (or meson
exchange currents) is larger than the errors in the data, even 
at low $Q^2$, and is not understood.

\item There is evidence that new physics (either in
the from of the $\rho\pi\gamma$ exchange current or
something else) is beginning to show up in the $A$
structure function above $Q^2$ of 2 - 3 GeV$^2$.    

\item The deuteron form factors provide
no evidence for the onset of pQCD, but quark cluster
models could explain the data. 

\end{itemize}
Study of the experimental situation leads to the
following conclusions:

\begin{itemize}  

\item The minimum of $B$ is very sensitive to details of 
the models, and improved measurements of $B$
for $Q^2$ in the region 1.5 - 4 GeV$^2$ are particularly
compelling.  It is important to accurately map out the
zero in the $B$ structure function.

\item Detailed disagreements between theories and
different data sets suggests the need for precision
studies at low $Q^2$.
  
\end{itemize}

\section{Threshold electrodisintegration}

\begin{figure}
\caption{Left panel: The cross section for threshold
electrodisintegration  of the deuteron.  (See Ref.~[1] for
discussion of the data and the theory.)  Right panel: 
The variation of $W^2$ with the photon energy
$\nu$ for various values of $x$.
The shaded regions show the approximate thresholds for the
production of bands of nucleon resonances.  The numbers in the
small circles are the number of distinct channels in each band.}
\leftline{\includegraphics[height=.35\textheight]{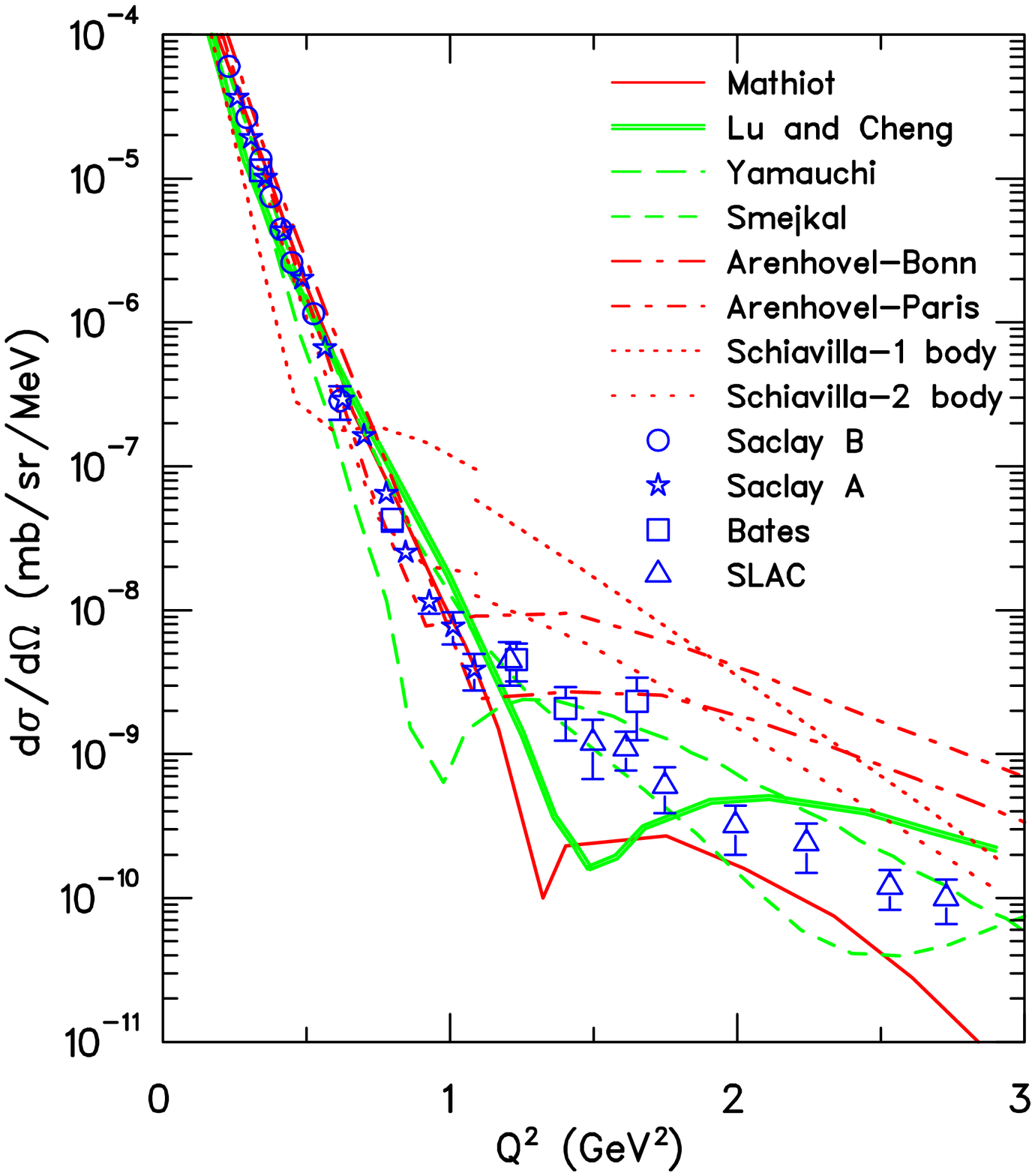}
\hspace{0.4in}
\includegraphics[height=.3\textheight]{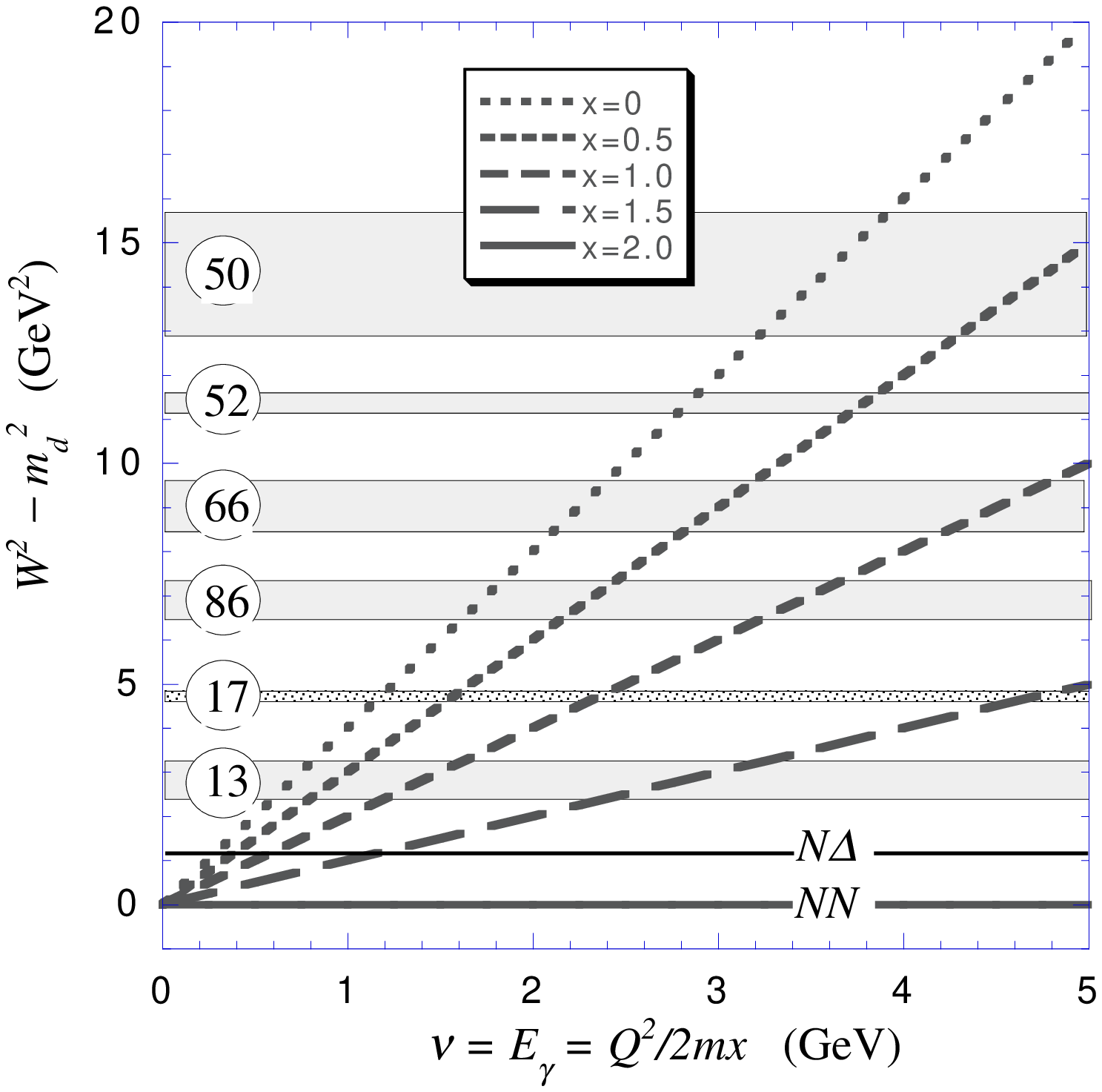}}
\label{thresh}
\end{figure}

Threshold deuteron electrodisintegration measures the $d(e,e^{\prime}) 
pn$ reaction in kinematics in which the proton and neutron, rather than 
remaining bound, are unbound with a few MeV of relative kinetic energy 
in their center of mass system.  If the final state energy is low 
enough, the final state will be dominated by transitions to the $^1S_0$ 
final state, and will be a pure $\Delta S = 1$, $\Delta I = 1$, $M1$  
transition, similar to the $N\to\Delta(1232)$ transition.  This 
transition is a companion to the $B$ structure function; both are 
magnetic transitions and both are filters for exchange currents with 
only one isospin ($d\to d$ is $\Delta I=0$ and $d\to\ ^1S_0$ is $\Delta 
I=1$).  To see the similarity, compare the top right panel of 
Fig.~\ref{fig:RABT20} with the threshold measurements shown in left 
panel of Fig.~\ref{thresh}.  Both have a similar shape, and in both 
cases the uncertainties in the theoretical predictions are large.  

The similarity of these two processes (elastic and threshold inelastic) 
also holds for the theory.  These two processes can be used to 
separately determine the precise details of the $I=0$ and $I=1$ 
exchange currents.   Once the exchange currents are fixed, they can be 
used to predict the results of $d(e,e'p)n$ over a wide kinematic 
region.  Any theoretical approach that works for the form factors 
should also work equally well for threshold electrodisintegration, yet 
very few of the groups who have calculated form factors have also 
calculated the threshold process.  These calculations, when completed, 
will provide a more definitive test of the various relativistic 
approaches discussed in the previous sections.

\section{Deuteron photodisintegration}

Since {\it both\/} the recent deuteron form factor measurements {\it 
and\/} the recent high energy deuteron photodisintegration measurements 
have been made with 4 GeV electron beams, it is sometimes assumed that 
the same theory should work for both.  This need not be the case, 
because the kinematics of elastic electron-deuteron scattering and 
deuteron photodisintegration are very different, and the physics being 
explored by these two measurements is also very different.  The 
implications of this remarkable feature of electronuclear physics is 
often not fully appreciated.

The kinematics of elastic scattering and photodisintegration are 
compared in the right panel of Fig.~\ref{thresh}, which shows 
$W^2-m_d^2$ as a function of the photon (real or virtual) energy.  Here 
$x=Q^2/(2m\nu)$ is the familiar Bjorken scaling variable.  The mass of 
the final excited state increases rapidly as $x$ decreases below its 
maximum allowed value of $x=m_d/m\simeq2$.   For any energy $\nu$ or 
any $Q^2$, elastic $ed$ scattering leaves the $pn$ system bound, with 
no internal excitation energy added to the two nucleons.  As $x$ 
$\rightarrow$ 0 (the real photon limit) the maximum value of $W$ is 
reached for any given beam energy.   The 24 well established nucleon 
resonances,  and the bands of thresholds at which these resonances are 
excited, are show in Fig.~\ref{thresh}.   At $E_{\gamma}$ = 4 GeV, the 
final state mass is approximately 4.5 GeV, and at least 286 thresholds 
for the production of pairs of baryon resonances have been crossed (and 
there are probably more from unseen or weakly established resonances).  
A photon energy of 4 GeV corresponds to $np$ scattering with an $np$ 
laboratory kinetic energy of about 8 GeV!

It is clearly very difficult (if not impossible) to construct a theory 
of high energy photoproduction in which all of these resonances and 
their corresponding 286 production thresholds are treated 
microscopically.  By contrast, elastic electron deuteron scattering 
requires a microscopic treatment of only {\it one channel\/}.  All of 
the 286 channels also contribute to elastic scattering, of course, but 
in this case they are {\it not explicitly excited\/}, and can probably 
be well described by slowly varying short-range terms included in a 
meson exchange (or potential) model.  In photodisintegration, {\it each 
of these channels is excited explicitly\/} and an alternate framework 
that {\it averages over the effects of many hadronic states\/} is 
needed.  The alternatives are to use a Glauber-like approach, or to 
borrow from our knowledge of DIS and build models that rely on the 
underlying quark degrees of freedom.   

\subsection{High Energy Photodisintegration}

\begin{figure}
\caption{Photodisintegration cross 
section
$s^{11}d\sigma/dt$ versus incident lab photon energy.
The calculations are from 
Kang, Erbs, Pfeil and Rollnik (solid line),
Lee (dashed line),
Raydushkin quark exchange (dot-dashed line),
reduced nuclear amplitudes of Brodsky Hiller (dotted line),
quark qluon string model (short dashed line), and
Frankfurt, Miller, Strikman and Sargsian QCD
rescattering (shaded region).  (See Ref.~[1] for references
and discussion of the data and theory.)}
\includegraphics[height=.68\textheight]{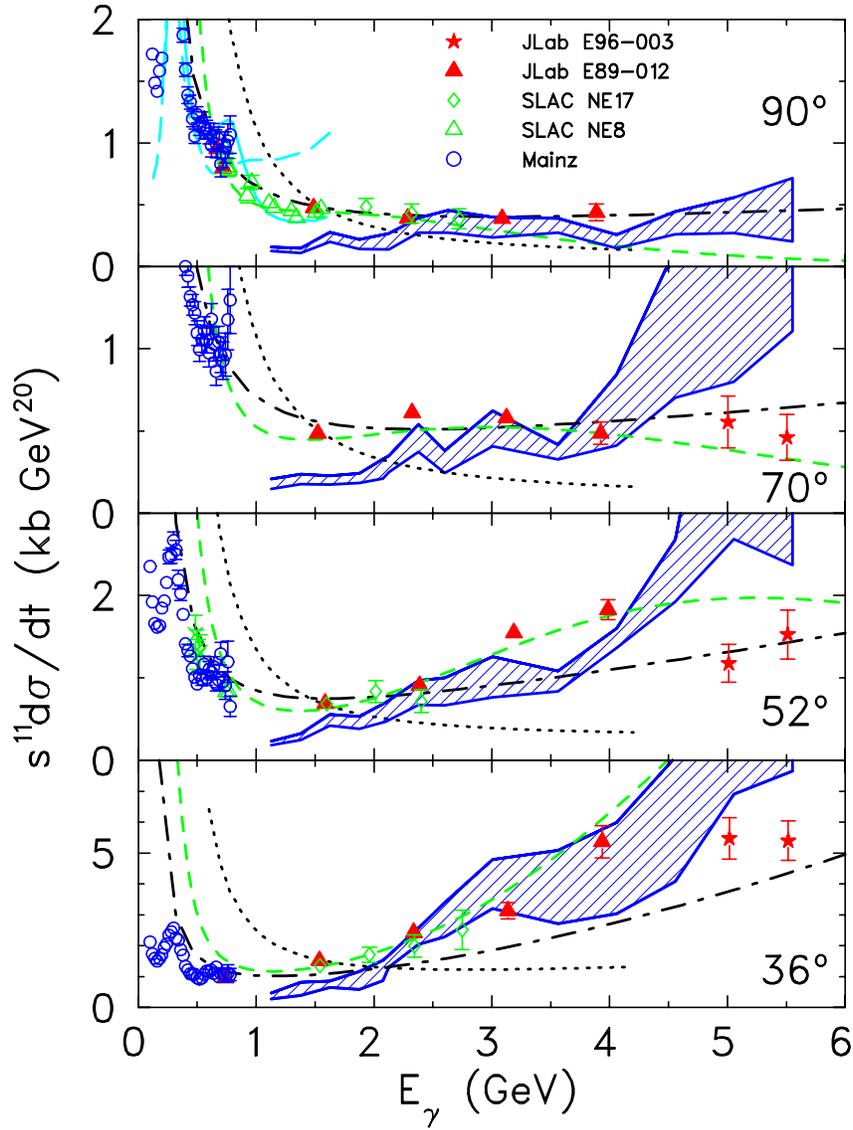}
\label{fig:dgpsigmas}
\end{figure}
  
Figure \ref{fig:dgpsigmas} shows the published high energy 
photodisintegration data, from experiments NE8 
\cite{napolitano88,freedman93} and NE17 \cite{belz95} at SLAC, and 
E89-012 \cite{bochna98} and E96-003 \cite{schulte01}  at CEBAF. These 
experiments determine cross sections for $\theta_{\rm cm}$ $\approx$ 
36$^{\circ}$, 52$^{\circ}$, 69$^{\circ}$,and 89$^{\circ}$ at energies 
from about 0.7 to 5.5 GeV; there are also some backward angle data up 
to 1.6 GeV from NE8.

The main feature of the data above about 1 GeV is the $s^{-11}$ 
($s^{-10}$) fall off (where $s=(p_1+p_2)^2$ is the square of the cm.\ 
energy)  of the cross sections $d\sigma/dt$  ($d\sigma/d\Omega$) at 
$\theta_{\rm cm}$ = 90 and 69$^{\circ}$, in  agreement with 
perturbative QCD expectations. In contrast, the cross sections at the 
forward angles 36 and 52$^{\circ}$, fall off more slowly, with $\approx 
s^{-9}$ scaling at lower energies, until the onset of the $s^{-11}$ 
behavior at about 4 and 3 GeV beam energy, respectively. At each angle 
the onset of the $s^{-11}$ behavior corresponds to a perpendicular 
momentum, $p_T$, of approximately 1 GeV$^2$. 

\begin{figure}
\caption{Left panel: Induced polarization for deuteron 
photodisintegration at $\theta_{\rm cm}$ $=$ 90$^\circ$.
The calculations are from Kang, Erbs, Pfeil and Rollnik
(dashed line)  and from Sargsian (dot-dashed line).  Right
panel: Polarization transfer for 
deuteron photodisintegration at $\theta_{\rm cm}$ $=$ 90$^\circ$.
The calculation is from Schwamb and Arenh\"ovel.}
\leftline{\includegraphics[height=.3\textheight]
{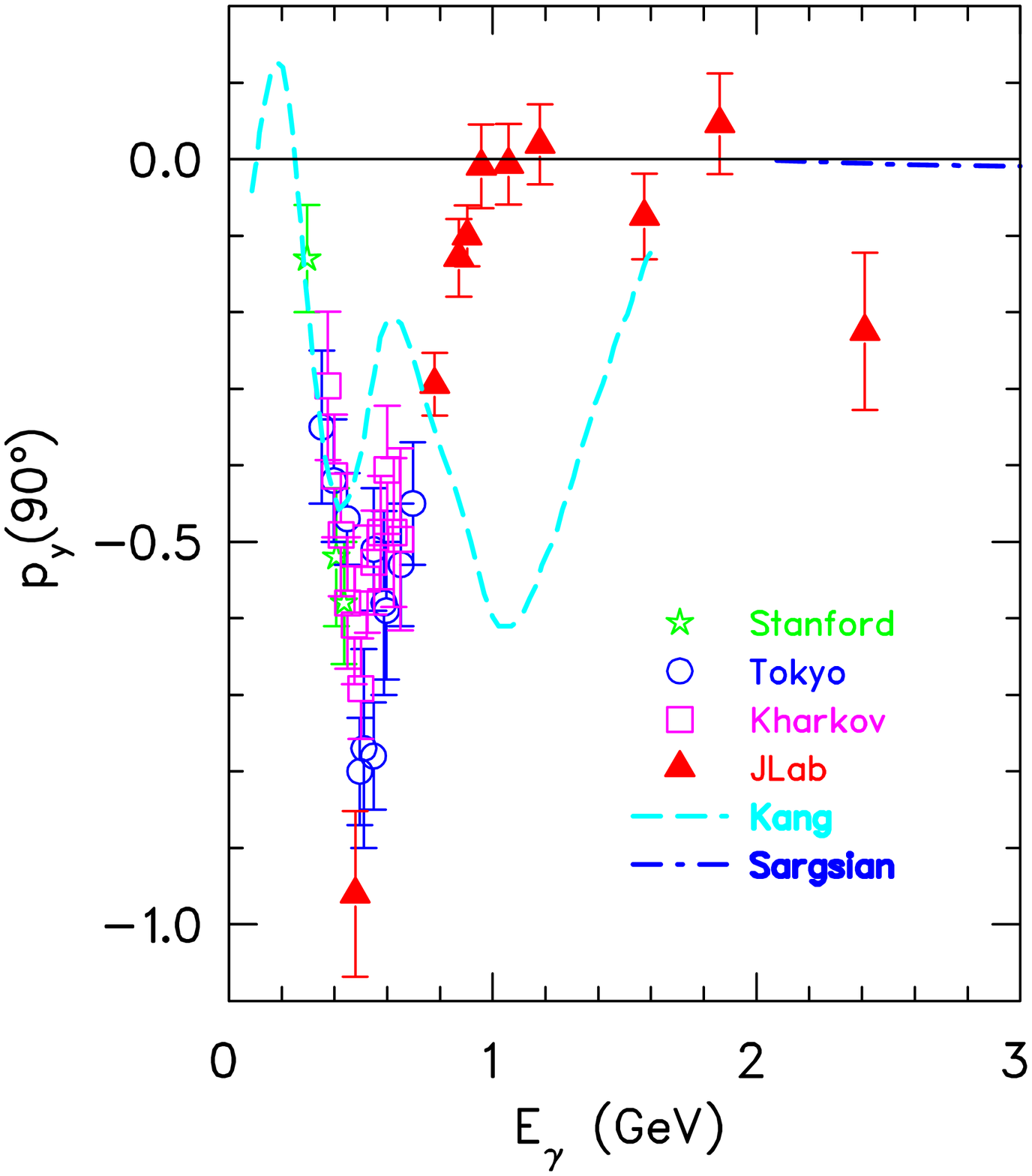}
\hspace{0.7in}\includegraphics[height=.3\textheight]{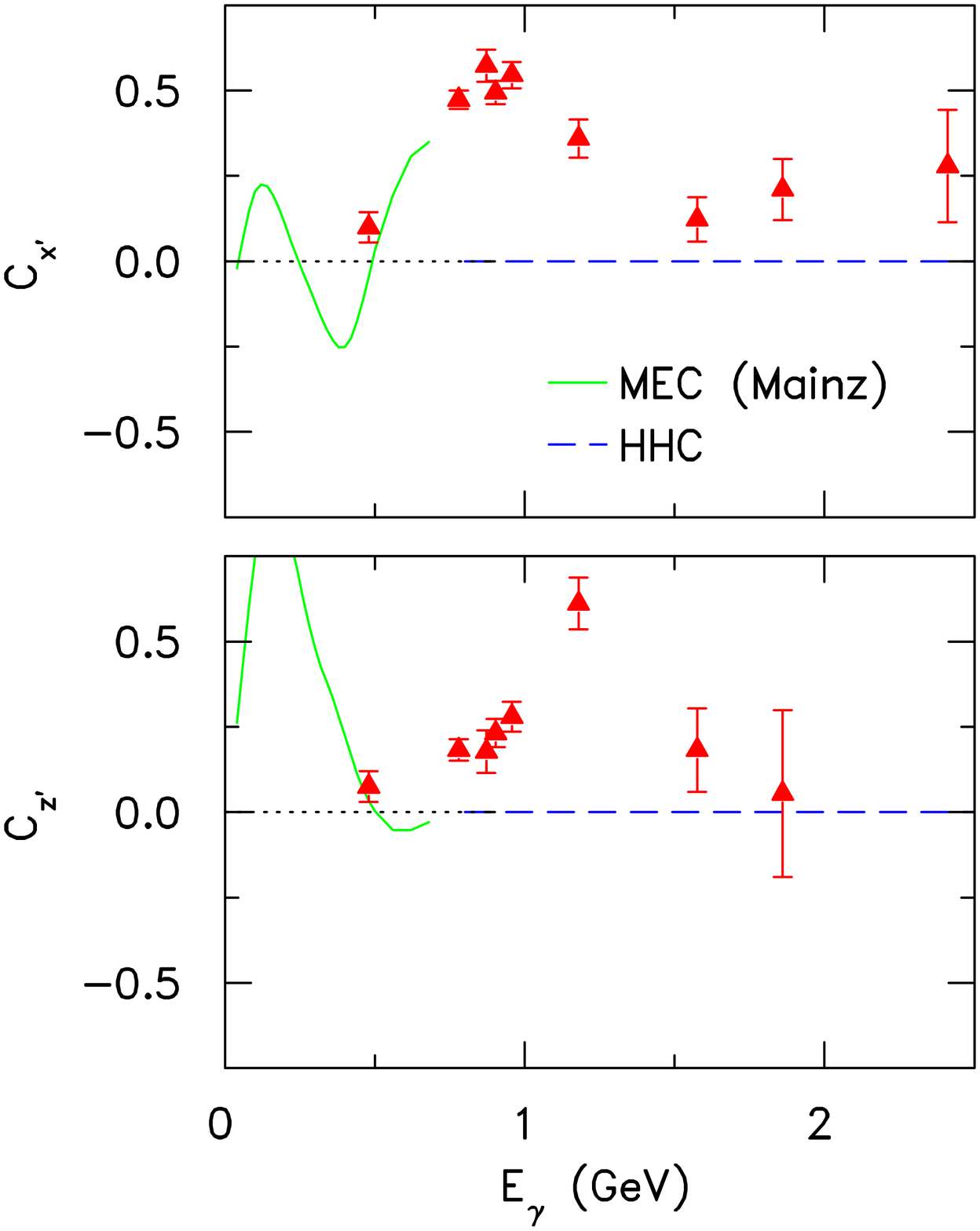}}
\label{fig:89019py}
\end{figure}

The highest energy polarization measurements are of $p_y$ (the induced 
polarization of the proton), and $C_{x'}$ and $C_{z'}$ (the transfer of 
circular polarization from the photon to the proton) from CEBAF E89-019 
\cite{wije01}. The left panel of Fig.~\ref{fig:89019py} shows the 
striking feature that  the induced polarization $p_y$   is consistent 
with zero (as predicted by pQCD) at energies above about 1 GeV, the 
same energy at which the $s^{-11}$  cross section scaling begins. The 
right panel of the same figure shows that the polarization transfer  
observables both appear to peak near 1 GeV, and decrease at  higher 
energies.

\subsection{Conclusions (deuteron 
photodisintegration)}

Review of deuteron photodisintegration suggests the following:

\begin{itemize} 

\item A microscopic meson-baryon theory of deuteron
photodisintegration must describe the $NN$
interaction at high energies, including pion production  and
the  contributions of hundreds of $N^*$ channels. It is
unlikely that such a theory will be constructed in the
foreseeable future. 

\item For $p_T^2> 1$ GeV$^2$, cross sections appear to
follow the constituent counting rules, but it is 
expected that an absolute pQCD calculation would greatly
underpredict the data.
Similar observations may be made for other photoreactions,
and it remains to be seen how this behavior arises,
and if there is a general explanation for it.

\item Some nonperturbative quark models do well describing the
data qualitatively. Further theoretical development and 
experimental tests of these models would be
desirable.

\end{itemize}

\section{Overall Conclusions}

Our overall conclusions from the study of form factors and high
energy photodisintegration can be briefly summarized as follows:

\begin{itemize}

\item Meson theory works well in cases where all of the {\it
active\/} hadronic channels that can contribute to a
process are included.  This has been done for the deuteron form
factors (where only the $NN$ channel is active), but is
impossible for high energy deuteron photodisintegration where
100's of $N^*N^*$ channels are active.  At high energy, any
successful meson theory must include relativistic effects.

\item New approaches, probably using quark degrees of freedom, 
are needed for high energy deuteron photodisintegration.

\item Meson theory behaves as might be expected.
It works for the deuteron form factors and
does not work for photodisintegration (because of the number of
channels).
 
\item Perturbative QCD does not behave as might be expected. 
Theoretically it should work (or not work)
equally well for the deuteron form factors at high $Q^2$ as it
does for photodisintegration at high $p_\perp^2$ (both are
exclusive processes).   However the scaling laws seem to work
qualitatively for photodisintegration and to fail badly for
elastic scattering.

\end{itemize}

\noindent Please refer to Ref.~[1] for a more complete
discussion of all of the points in this talk.

\begin{theacknowledgments}
This work was supported in part by the US Department
of Energy. The Southeastern 
Universities Research Association (SURA) operates the Thomas
Jefferson National Accelerator Facility under DOE contract
DE-AC05-84ER40150.  
\end{theacknowledgments}

\end{document}